\date{November 1, 2010}
\documentclass{article} 
\usepackage{amsthm}

\begin{document}

\title{A note comprising a negative resolution of the Efficient Market Hypothesis\footnote{We are grateful to Volker Nannen for valuable comments.}}
\author{Robert Viragh\footnote{Email: rviragh@gmail.com}}
\maketitle

\begin{abstract}
This note comprises a negative resolution of the Efficient Market Hypothesis.
\end{abstract}

\noindent
\textbf{Keywords:} Efficient Market Hypothesis, EMH, Efficient Market Model

\section{Resolution}
The Efficient Market Hypothesis states that all consequences of available information are factored into a security's price.~\cite{Fama}

To resolve the hypothesis, follow the methodology below, keeping this methodology both public and credible (i.e. through legal instruments):
\begin{enumerate}
\item Create a fund with \$1 billion, to be paid out after 300 trading days.
\item Select 1000 micro-cap stocks, each trading for under \$50 million in market capitalization, enumerate them 0--999 (their ``codes''), publishing this list.
\item By means of a secure computer:
\begin{itemize}
\item[-] Pick a number, $W$ (for winner) in the range 0--999 randomly, securely.  The company with code $W$ will be paid by the fund.  Keep $W$ confidential.
\item[-] Randomly, securely, find a prime, $P$, of 300 digits such that:
\begin{equation}floor( (P \bmod 10000) / 10) = W\end{equation}
In other words, the first three of its last four digits equal $W$. (For example if $W=5$ the prime could end $...0057$, if $W=999$, it could end $...9993$).  Keep $P$ confidential except as described below.
\item[-] Randomly, securely, find another, smaller, prime, of 300 digits.  Keep this prime confidential.
\item[-] Multiply the two primes, producing a large number.
\item[-] Publish this large number.  Equation (1) is public as well.
\end{itemize}
\item Observe trading for 300 trading days, performing the following action at the end of each trading day with the help of the secure computer:
\begin{itemize}
\item[-] After trading closes, reveal one digit of $P$: its most significant digit on the first day, then the next-most-significant digit each day thereafter.
\end{itemize}
\item After revealing all of the digits, pay out the \$1 billion to company with code $W$, as promised and bound by legal instruments.
\end{enumerate}
At what point does the chosen company's stock reflect the fact that it is receiving \$1 billion?  Obviously on the last four nights, the winning code has been revealed in the plain, and since the methodology is credible the pricing of company with code $W$ must surely reflect the imminent cash influx.  But a few nights before, the market has been given nearly all of $P$, so that there are just 10,000 or 100,000 possibilities to test, and, by the Fundamental Theorem of Arithmetic, the market can be sure when it has found the correct factor.  Testing a few thousand possible factors is obviously a trivial operation.

Therefore, the market cannot be so dumb that it does not reflect the imminent transfer in the price of stock $W$ near the end of the 300 days.  However, just as obviously, the market is not so smart that it can factor a 600-digit number, a prohibitively difficult task forming the basis of secure cryptographic protocols such as RSA (see \cite{Cormen}), overnight either.  Therefore, the result of the experiment will clearly show market inefficiency\footnote{We note that efficiency with respect to the price of computing, beyond presupposing a market that is not efficient by the common definition of the term, is not possible either, because there is no such ``price'' --- it is different for Intel, Wall Street, etc. --- just as there is no ``price'' for an analysis equal to Warren Buffett's on a given question.}: the market prices cannot reflect the completed analysis until some time between the first and last days.

Until that time, its prices will not reflect all of the available information, and, therefore, the Efficient Market Hypothesis has been resolved negatively.\hfill{}\qedsymbol{} 
\par\vspace{5mm}
Our results should not be overstated.  Although the market cannot be efficient, that does not mean it is not very clever, and we would not try to beat it on an empty stomach.

\end{document}